\vskip10pt
\centerline {\bf Quantum number conservation in intergenerational interactions}
\vskip5pt
\centerline {Douglas Newman}
\vskip5pt
\centerline {e-mail: \it dougnewman276@gmail.com}
\vskip20pt

\beginsection Abstract

The seven binary quantum numbers that distinguish fundamental fermions
have been shown to be conserved in decays and interactions.
Here applications of this law are clarified to take account of odd (uct) 
and even (dsb) parity quarks defining separate representations of SU(3), 
each with its own definition of the F and G quantum numbers that 
distinguish generations. These representations are related by the 
CKM unitary matrix. The SU(3) groups define 
an SU(6) $\equiv$  SU(3)$\otimes$U(1)$\otimes$SU(3) group of
transformations of all six quarks. Quark/anti-quark structures of
J=0 mesons are shown to correspond to all the SU(6) generators. 
Applications of quantum number conservation to fermion and meson
interactions, which take account of the CKM matrix, are described.

\vskip 40pt

\beginsection \S1 Introduction

Clifford unification (CU) [1,2] provides a description of all the
observed fundamental fermions in terms of seven commuting elements of the 
Clifford algebra $Cl_{7,7}$. Their binary $(\pm1)$ quantum 
numbers, labelled A,...,G, distinguish all the different fermions, 
where F,G distinguish fermion generations.
It was reported in [2] that fermion interactions and decays 
are subject to the conservation of all seven quantum numbers. 
 
The quantum numbers A,B,C describe the way in which the fermions fit
into space-time. `B' distinguishes fermions from anti-fermions, and has the 
effect of counting the number of fermions in composites. `A' determines spin
directions in an arbitrary orientation. These quantum numbers are defined 
by the 4-spinors in Dirac's theory of the electron (e.g. see \S4.6 of [5]).
The parity quantum number C is defined 
by the eigenvalues of a Lorentz invariant element of the $Cl(3,3)$
algebra determined in \S2 of [2]. C also describes
fermion intrinsic parities, which distinguish the 
spatial coordinates employed in transformations
of their (Dirac) spinors. This work is concerned with the dependence of the F,G quantum numbers 
on  the parity quantum number C=$\pm 1$ which, as noted in [3], 
was not taken into account in [2]. In CU all the fundamental fermions 
are members of doublets with opposite parities. 

High energy experiments provide detailed information on the
decays and interactions of charged particles. It was shown in [2] 
that observed fermion decays and interactions satisfy
quantum number conservation (QNC),  
providing a useful tool in the design of experiments and interpretation 
of their results. The effects of the Cabibbo-Kobayashi-Maskawa (CKM) matrix,
which are the subject of this work, provide a modification of QNC.
 
The SU(3) (F,G) labelling of fermion generations is analogous
with the SU(3) quark colour (D,E) quantum number labelling, and replaces 
the original SU(3) flavour labels, described, for example, in \S9.6 of [5]. It
is shown that quark generations can be described by two SU(3) subgroups that are
united into an SU(6)$\equiv$SU(3)$\otimes$U(1)$\otimes$SU(3) group of
transformations (often referred to as 331). The mathematical analysis has been 
assisted by the detailed study of the 331 group structure by Hartanto and Handoko [6]. 
Existing physical interpretations of this and other SU(6) structures [7-10], are concerned
with developing Lie group Grand Unification schemes and bear no relation to the 
present work, which is based on the existing $Cl_{7,7}$ algebraic structure
of Clifford Unification [1,2].

\beginsection \S2 SU(3) structure of the three observed quark generations

Clifford unification distinguishes u,c,t quarks, which have negative intrinsic parity
with quantum number C=${-1}$, and d,s,b quarks which have positive intrinsic parity
with quantum number C=${+1}$. The C=${-1}$ and C=${+1}$ quarks each define three 
dimensional SU(3) vector bases 
(here denoted SU(3)$_-$ and SU(3)$_+$ respectively). Quark colour distinctions are not 
relevant in the description of generations. The following discussion focusses on the
SU(3)$_-$ properties of u,c,t quarks, but the mathematical
description of SU(3)$_+$ d,s,b quarks is the same. C=${-1}$ quarks are 
expressed as the three dimensional vectors
$$
\rm u\equiv \left(\matrix{1 \cr 0 \cr 0}\right),\>\rm c \equiv \left(\matrix{0 \cr 1 \cr 0}\right),
\>\rm t \equiv \left(\matrix{0 \cr 0 \cr 1}\right),                                     \eqno(2.1)
$$
and the corresponding antiquarks as the transposed vectors 
$$
\rm \bar u\equiv\left(\matrix{1 & 0 & 0}\right),\>\rm \bar c \equiv \left(\matrix{0 & 1 & 0}\right),
\>\rm \bar t\equiv\left(\matrix{0 & 0 & 1}\right).                                     \eqno(2.2)
$$
J=0 mesons are represented by $3\times3$ matrices with a single non-zero 
element, where the accepted meson identifications of
quark/anti-quark structures are given in Table A.2 of the Appendix. For example
$$
\bar{\rm D}^0\equiv{\rm u\bar c}= \left(\matrix{0&1&0 \cr 0&0&0 \cr 0&0&0}\right),\>\>
{\rm D}^0    \equiv{\rm c\bar u}= \left(\matrix{0&0&0 \cr 1&0&0 \cr 0&0&0}\right). \eqno(2.3)
$$
The standard Gell-Mann $3\times3$ matrix representation of SU(3) generators 
$\lambda_i$ is employed in this section. 
The relationships of the $\lambda_i;{i=1,2,4,5,6,7}$ matrices with the 
interpretations of mesons in terms of their quark/anti-quark structures 
are given in Tables 2.1 and 2.2 below, where `effect' refers to the change in generation.
In the absence of any
spatial contribution, combinations of a C=1 quark with a C=$-1$ anti-quark, and
visa-versa, give meson parities $1\times -1 = -1$. Table 2.1 lists combinations
C=$-1$ quarks and C=1 anti-quarks giving mesons with C=$-1$.  Table 2.2 lists combinations
C=1 quarks and C=$-1$ anti-quarks giving, again, mesons with C=$-1$.
$$\vbox 
{\settabs 8 \columns\+  Table 2.1: F,G quantum numbers of SU(3)$_-$ generating mesons \cr  
	\+|||||||||||||||||||||||||||||||||\cr
	\+q$\bar{\rm q}$ &F &G &$\>\>\>\>$ matrix& $\>\>\>\>\>$ effect&meson\cr
	\+|||||||||||||||||||||||||||||||||\cr
	\+c$\bar{\rm u}$&$\>\>0$  &$-2$      &${(\lambda_{1}-i\lambda_2)}/2$ &$\>\>\>\>\>1\to2$&$\>\>$D$^0$\cr
	\+u$\bar{\rm c}$&$\>\>0$  &$\>\>2$   &${(\lambda_{1}+i\lambda_2)/2}$ &$\>\>\>\>\>2\to1$&$\>\>\bar{\rm D}^0$ \cr 
	\+t$\bar{\rm u}$&$-2$     &$\>\>0$   &${(\lambda_4-i\lambda_5)/2}$   &$\>\>\>\>\>1\to3$&$\>\>$T$^0$         \cr
	\+u$\bar{\rm t}$&$\>\>2$  &$\>\>0$   &${(\lambda_4+i\lambda_5)/2}$   &$\>\>\>\>\>3\to1$&$\>\>\bar{\rm T}^0$ \cr
	\+t$\bar{\rm c}$&$-2$     &$\>\>2$   &${(\lambda_6-i\lambda_7)/2}$   &$\>\>\>\>\>2\to3$&$\>\>$T$^0_c$       \cr
	\+c$\bar{\rm t}$&$\>\>2$  &$-2$      &${(\lambda_6+i\lambda_7)/2}$   &$\>\>\>\>\>3\to2$&$\>\>\bar{\rm T}^0_c$\cr
	\+|||||||||||||||||||||||||||||||||\cr}
$$

$$\vbox 
{\settabs 8 \columns\+  Table 2.2: F,G quantum numbers of SU(3)$_+$ generating mesons \cr
	\+|||||||||||||||||||||||||||||||||\cr
	\+q$\bar{\rm q}$ &F &G &$\>\>\>\>$ matrix& $\>\>\>\>\>$ effect&meson\cr
	\+|||||||||||||||||||||||||||||||||\cr
	\+s$\bar{\rm d}$&$\>\>0$  &$-2$      &${(\lambda_1+i\lambda_2)/2}$ &$\>\>\>\>\>1\to2$&$\>\>$$\bar{\rm K}^0$\cr 
	\+d$\bar{\rm s}$&$\>\>0$  &$\>\>2$   &${(\lambda_1-i\lambda_2)/2}$ &$\>\>\>\>\>2\to1$&$\>\>$K$^0$\cr   
	\+b$\bar{\rm d}$&$-2$     &$\>\>0$   &${(\lambda_4-i\lambda_5)/2}$ &$\>\>\>\>\>1\to3$&$\>\>\bar{\rm B}^0$ \cr 	
	\+d$\bar{\rm b}$&$\>\>2$  &$\>\>0$   &${(\lambda_4+i\lambda_5)/2}$ &$\>\>\>\>\>3\to1$&$\>\>$B$^0$    \cr		
	\+b$\bar{\rm s}$&$-2$     &$\>\>2$   &${(\lambda_6-i\lambda_7)/2}$ &$\>\>\>\>\>2\to3$&$\>\>\bar{\rm B}^0_s$\cr
	\+s$\bar{\rm b}$&$\>\>2$  &$-2$      &${(\lambda_6+i\lambda_7)/2}$ &$\>\>\>\>\>3\to2$&$\>\>$B$^0_s$   \cr 	    
	\+|||||||||||||||||||||||||||||||||\cr}
$$
Identifications of neutral mesons with {\it combinations}
of the SU(3) generators $\lambda_i$ are different to the identification of gluons, which
correspond to the $\lambda_i$ as mediators of colour exchange. This is relevant because
of the ongoing discussion of the experimental evidence that the physical K mesons are
not described by $\bar{\rm K}^0\equiv {\rm s}\bar{\rm d}$ and  
${\rm K}^0\equiv {\rm d}\bar{\rm s}$ (e.g. [5]\S14.4). The G=0 quantum number 
provides further evidence of this, in that it makes possible the decay 
${\rm s}\bar{\rm d}+{\rm d}\bar{\rm s}\to 2\pi^0$. Analogous problems arise with
the $\bar{\rm B}^0$ and ${\rm B}^0$ mesons (e.g. [5]\S14.6).

The Gell-Mann $3\times3$ diagonal matrix representations of the generators $\lambda_3$ 
and $\lambda_8$ for SU(3) are
$$
\lambda_3 = \left(\matrix{1 &0&0\cr 0 &-1& 0\cr 0 &0& 0\cr}\right),\>\>\>\>
\lambda_8 = {1\over \sqrt 3}\left(\matrix{1 &0&0\cr 0 &1& 0\cr 0&0&-2\cr}\right).       \eqno(2.4)
$$
In the case of SU(3)$_-$, these matrices correspond to the quark compositions   
$$
\rm\lambda_3= (u\bar u - c\bar c), \>\>\>\>\>
\rm \lambda_8={1\over {\sqrt 3}}(u\bar u + c\bar c - 2t\bar t).         \eqno(2.5)                   
$$
Analogous compositions for SU(3)$_+$ are
$$
\rm\lambda_3=(d\bar d - s\bar s), \>\>\>\>
\rm \lambda_8={1\over {\sqrt 3}}(d\bar d + s\bar s - 2b\bar b).         \eqno(2.6)                   
$$
These matrices do not correspond to observed particles 
as no possible identifications seem to have been reported in the literature. 

\beginsection \S3. SU(6) structure of intergenerational interactions

The six u,d,s,c,b and t quarks together provide the basis of SU(6) 
in which the generation changing interactions are the 35 
J=0 mesons. These comprise twelve neutral (C=0) mesons listed in Tables 2.1 
and 2.2, five C=0 particles constructed from undefined traceless combinations
of $\rm u\bar u$, $\rm c\bar c$, $\rm t\bar t$, $\rm d\bar d$,
$\rm s\bar s$ and $\rm b\bar b$ discussed above and
the 18 C=$\pm 2$ mesons listed in Table 3.1 (below). Meson/anti-meson 
mixing cannot occur in the case of charged mesons.

Mathematical expressions relating to the construction of SU(6) from two SU(3) groups
have been obtained by Hartanto and Handoko [6]. In particular they provide 
(in their Appendix) a list of the matrix representations of all 35 generators of SU(6).
These are related to the SU(3) labels as follows:
\vskip1pt
\item{SU(3)$_+$:} The labelling in Table 2.1 is retained
\vskip2pt
\item{SU(3)$_-$:} The labelling in Table 2.2 is changed to that in [6], with $\lambda_i\to \lambda_{(i+26)}$
\vskip2pt
\item{SU(6)$\>\>\>$:} Labels of the remaining 18 C=1 off-diagonal elements of SU(6) are given in Table 3.1 below.
\vskip2pt
$$\vbox 
{\settabs 9 \columns\+  Table 3.1: CFG signatures of the J=0
	mesons with standard identifications\cr  
	\+&||||||||||||||||||||||||||||||\cr
	\+&q$\bar{\rm q}$ &F &G &$\>\>\>\>\>$labelling &in [6] &meson\cr
	\+&||||||||||||||||||||||||||||||\cr
	\+&d$\bar{\rm u}$&$\>\>0$  &$\>\>0$   &$\>\>\>\>\>\>\>\>\>\>\>\>\>{1\over 2} (\lambda_{9}$&$-i\lambda_{10})$&$\>\>\>\pi^-$\cr 
	\+&s$\bar{\rm c}$&$\>\>0$  &$\>\>0$   &$\>\>\>\>\>\>\>\>\>\>\>\>\>{1\over 2} (\lambda_{17}$&$\>-\>\>i\lambda_{18})$&$\>\>$ D$^-_s$\cr 
	\+&b$\bar{\rm t}$&$\>\>0$  &$\>\>0$   &$\>\>\>\>\>\>\>\>\>\>\>\>\>{1\over 2}
	(\lambda_{25}$&$\>-\>\>i\lambda_{26})$&$\>\>$ T$^-_b$\cr
	\+&s$\bar{\rm u}$&$\>\>0$  &$-2$      &$\>\>\>\>\>\>\>\>\>\>\>\>\>{1\over 2} (\lambda_{15}$&$\>-\>\>i\lambda_{16})$&$\>\>\>\>$K$^-$\cr	
	\+&d$\bar{\rm c}$&$\>\>0$  &$\>\>2$ &$\>\>\>\>\>\>\>\>\>\>\>\>\>{1\over 2} (\lambda_{11}$&$\>-\>\>i\lambda_{12})$&$\>\>\>\>$D$^-$\cr
	\+&b$\bar{\rm u}$&$-2$    &$\>\>0$  &$\>\>\>\>\>\>\>\>\>\>\>\>\>{1\over 2}
	(\lambda_{21}$&$\>-\>\>i\lambda_{22})$&$\>\>$ B$^-$\cr 
	\+&d$\bar{\rm t}$&$\>\>2$&$\>\>0$ &$\>\>\>\>\>\>\>\>\>\>\>\>\>{1\over 2}
	(\lambda_{13}$&$\>-\>\>i\lambda_{14})$&$\>\>$ T$^-$\cr
	\+&s$\bar{\rm t}$&$\>\>2  $&$-2$ &$\>\>\>\>\>\>\>\>\>\>\>\>\>{1\over 2}
	(\lambda_{19}$&$\>-\>\>i\lambda_{20})$&$\>\>$ T$^-_s$\cr 	
	\+&b$\bar{\rm c}$&$-2$ &$\>\>2$ &$\>\>\>\>\>\>\>\>\>\>\>\>\>{1\over 2}
	(\lambda_{23}$&$\>-\>\>i\lambda_{24})$&$\>\>$ B$^-_c$ \cr	
	\+&&&&&&&\cr
	\+&u$\bar{\rm d}$   &$\>\>0$   &$\>\>0$&$\>\>\>\>\>\>\>\>\>\>\>\>\>{1\over 2}	
	(\lambda_{9}$&$\>+\>\>i\lambda_{10})$&$\>\>\>\pi^+$\cr
	\+&c$\bar{\rm s}$   &$\>\>0$   &$\>\>0$&$\>\>\>\>\>\>\>\>\>\>\>\>\>{1\over 2} (\lambda_{11}$&$\>+\>\>i\lambda_{12})$&$\>\>$ D$^+_s$\cr		
	\+&t$\bar{\rm b}$  &$\>\>0$   &$\>\>0$&$\>\>\>\>\>\>\>\>\>\>\>\>\>{1\over 2}
	(\lambda_{25}$&$\>+\>\>i\lambda_{26})$&$\>\>$ T$^+_b$\cr 		 
	\+&u$\bar{\rm s}$   &$\>\>0$   &$\>\>2$ &$\>\>\>\>\>\>\>\>\>\>\>\>\>{1\over 2}	
	(\lambda_{15}$&$\>+\>\>i\lambda_{16})$&$\>\>\>\>$K$^+$\cr
	\+&c$\bar{\rm d}$   &$\>\>0$   &$-2$   &$\>\>\>\>\>\>\>\>\>\>\>\>\>{1\over 2}
	(\lambda_{11}$&$\>+\>\>i\lambda_{12})$&$\>\>\>\>$D$^+$\cr  
	\+&u$\bar{\rm b}$   &$\>\>2$  &$\>\>0$ &$\>\>\>\>\>\>\>\>\>\>\>\>\>{1\over 2} (\lambda_{21}$&$\>+\>\>i\lambda_{22})$&$\>\>$ B$^+$\cr 	 		    	           	
	\+&t$\bar{\rm d}$   &$-2$     &$\>\>0$ &$\>\>\>\>\>\>\>\>\>\>\>\>\>{1\over 2} (\lambda_{13}$&$\>+\>\>i\lambda_{14})$&$\>\>$ T$^+$\cr	
	\+&t$\bar{\rm s}$   &$-2$     &$\>\>2$ &$\>\>\>\>\>\>\>\>\>\>\>\>\>{1\over 2} (\lambda_{19}$&$\>+\>\>i\lambda_{20})$&$\>\>$ T$^+_s $\cr
	\+&c$\bar{\rm b}$   &$\>\>2$  &$-2$    &$\>\>\>\>\>\>\>\>\>\>\>\>\>{1\over 2} (\lambda_{23}$&$\>+\>\>i\lambda_{24})$&$\>\>$ B$^+_c$\cr		    
	\+&||||||||||||||||||||||||||||||\cr}
$$
See Table A.2 in the Appendix for the distinctions between the quantum numbers F$_+$ and F$_-$, G$_+$ and G$_-$ in Tables 2.1, 2.2 and 3.1.

Lepton generations have the same SU(6) structure as the quarks.
The neutral C=$-1$ leptons (i.e. neutrinos) are analogous to the u,c,t quarks,
with the SU(3)$_-$ basis
$$
|\nu_e> \equiv \left(\matrix{1 \cr 0 \cr 0}\right),\>|\nu_\mu> \equiv \left(\matrix{0 \cr 1 \cr 0}\right),
\>|\nu_\tau> \equiv \left(\matrix{0 \cr 0 \cr 1}\right).               \eqno(3.1)                                     
$$

The charged leptons (i.e. electrons, muons and tauons) are analogous to the d,s,b quarks,
with the SU(3)$_+$ basis
$$
|\rm e^->\equiv \left(\matrix{1 \cr 0 \cr 0}\right),\>|\mu^-> \equiv \left(\matrix{0 \cr 1 \cr 0}\right),
\>|\tau^-> \equiv \left(\matrix{0 \cr 0 \cr 1}\right)                   \eqno(3.2)
$$
Corresponding anti-leptons have opposite signs for all quantum numbers and
are represented by transposed vectors. Quantum number descriptions of the 
three observed generations of fermions are given in Table A.1 of the Appendix.

\beginsection \S4. Physical interpretations

Comparison of Tables 2.1 and 2.2 show, for example, equalities in the quantum numbers 
for ${\rm K}^0$ and $\bar{\rm D}^0$. Similar pairing occurs for $\bar{\rm K}^0$ and 
${\rm D}^0$. Table A.2 in the Appendix shows this relationship to be more complicated, 
but remains of physical interest. Analogous quantum number relationships also
exist for the ${\rm B}^0$/${\rm T}^0$ and ${\rm B}^0_s$/${\rm T}^0_c$ pairs. The
physical significance of these relationships need investigation.
In Table 2.3 the equalities F$_+$=1, G$_+$=1 between quantum numbers 
$\rm d\bar u \equiv s\bar c \equiv t\bar b$ and F$_+=-1$,  G$_+=-1$
between quantum numbers $\rm u\bar d \equiv c\bar s \equiv b\bar t$
suggest mixing of the designation of meson generations.

Five generators of SU(6) are represented by zero trace diagonal
matrices that are combinations of u$\bar{\rm u}$, s$\bar{\rm s}$,
t$\bar{\rm t}$, d$\bar{\rm d}$, c$\bar{\rm c}$ and b$\bar{\rm b}$.
It is not clear what relations exist between observed particles
and linear expressions in terms of these generators. Most of the
possible combinations involving G2 and G3 fermions will be
unstable or, at least, will have extremely short lifetimes.
Comparative mass arguments may provide one way forward. This is suggested 
by the familiar argument that the similar masses of $\pi^0,\pi^+,\pi^-$, 
provide a justification for the identification of 
$\rm \pi^0 = {1\over \sqrt 2}(u\bar u -d \bar d)$ (e.g. see [5],\S9.6.2).
Similar arguments suggest that short-lived particles might be described
by $ {1\over \sqrt 2}\rm(s\bar s -c \bar c)$ and 
${1\over \sqrt 2}\rm(b\bar b -t \bar t)$. 
\vskip20pt

\beginsection \S5 The CKM transformation relating SU(3)$_+$ and SU(3)$_-$ quark descriptions

QNC applications are complicated
by the different physical significance of F,G quantum numbers
for the two (C=$\pm1$) components of fermion doublets. 
The CFG signatures of the three generations of C= $\pm1$ quarks are 
related by the $3\times 3$ Cabibbo-Kobayashi-Maskawa (CKM) matrix 
which, apart from signs of its components, has been determined experimentally.
This involves a reinterpretation of the physical significance of the CKM matrix, 
avoiding the conventional assumption that dsb quarks have distinct weak and
mass eigenstates. Nevertheless, its mathematical interpretation as the relation
between dsb and uct eigenstates is the same here as that given in the
textbooks, with the additional assumption that all its
components are real. A detailed account of the properties of the CKM 
matrix is given in \S7.9 of [11]. In particular, it is made clear that
this matrix can be expressed as an SU(3) coordinate transformation involving 
three rotation angles.

Standard expressions for the CKM matrix and its inverse are
$$
{\bf V}_{\rm CKM} = \left(\matrix {\rm V_{ud} & \rm V_{us}&\rm V_{ub}\cr 
	\rm V_{cd} &\rm V_{cs}& \rm V_{cb}\cr
	\rm V_{td} &\rm V_{ts}& \rm V_{tb}\cr}\right)               \eqno (5.1)
$$
and
$${
{\bf V}^{-1}_{\rm CKM} = \left(\matrix{\rm V_{du} &\rm V_{su}&\rm V_{ bu}\cr 
	\rm V_{dc} &\rm V_{sc}& \rm V_{bc}\cr
	\rm V_{dt} &\rm V_{st}& \rm V_{bt}\cr}\right)} .            \eqno (5.2)
$$

Equations (5.1) and (5.2) can also be expressed in terms of relationships between quarks, viz.
$$\eqalign{
\{{\rm u}:\bar1\>1\>1\}_-&=\rm V_{ud}\{{\rm d}:1\>1\>1\}_+ + V_{us}\{{\rm s}:1\>1\>\bar1\}_+ 
+ V_{ub} \{{\rm b}:1\>\bar1\>1\}_+\cr
	\{{\rm c}:\bar1\>1\>\bar1\}_-&=\rm V_{cd}\{{\rm d}:1\>1\>1\}_+ + V_{cs}\{{\rm s}:1\>1\>\bar1\}_+ 
+ V_{cb} \{{\rm b}:1\>\bar1\>1\}_+\cr
	\{{\rm t}:\bar 1\>\bar1\>1\}_-&=\rm V_{td}\{{\rm d}:1\>1\>1\}_+ + V_{ts}\{{\rm s}:1\>1\>\bar1\}_+ 
+ V_{tb}\{{\rm b}:1\>\bar1\>1\}_+\cr}
\eqno (5.3)
$$ 
and
$$\eqalign{
	\{{\rm d}:1\>1\>1\}_+&=\rm V_{du}\{{\rm u}:\bar1\>1\>1\}_- + V_{dc}\{{\rm c}:\bar1\>1\>\bar1\}_- 
	+ V_{dt} \{{\rm t}:\bar1\>\bar1\>1\}_-\cr
	\{{\rm s}:\bar1\>1\>\bar1\}_+&=\rm V_{su}\{{\rm u}:\bar1\>1\>1\}_- + V_{sc}\{{\rm c}:\bar1\>1\>\bar1\}_- 
	+ V_{st} \{{\rm t}:\bar1\>\bar1\>1\}_-\cr
	\{{\rm b}:\bar1\>\bar1\>1\}_+&=\rm V_{bu}\{{\rm u}:\bar1\>1\>1\}_- + V_{bc}\{{\rm c}:\bar1\>1\>\bar1\}_- 
	+ V_{bt}\{{\rm t}:\bar1\>\bar1\>1\}_-\cr}
\eqno (5.4)
$$ 
where (CFG)$_-$ are the quantum numbers for the uct quarks and (CFG)$_+$ are the quantum numbers for the dcb quarks (given in Table A.1 of the Appendix).

The assignment of quantum numbers to fundamental fermions remains the same as in [2].
However, QNC as formulated in [2] must be adapted to include the 
equations (5.3) and (5.4). Nevertheless, the
{\sl dominant\/} decay and interaction processes still correspond to the conservation laws 
obtained in [2] in the approximation that the diagonal elements of the CKM matrix are unity
and its off-diagonal elements are zero. 

The diagonal elements of ${\bf V}_{\rm CKM}$, corresponding to decay 
and interaction processes that leave generations unchanged, always dominate. 
This is demonstrated by the magnitudes of components of 
given in \S7.9 of [11], viz.
$$
\left(\matrix {\rm |V_{ud}| & \rm |V_{us}|&\rm |V_{ub}|\cr 
	\rm |V_{cd}| &\rm |V_{cs}|& \rm |V_{cb}|\cr
	\rm |V_{td}| &\rm |V_{ts}|& \rm |V_{tb}|\cr}\right)
=\left(\matrix {.971 & .227 &.004\cr 
	.227&.973& .042\cr
	.008 &.042& .999\cr}\right)            \eqno (5.5)
$$
Although all the elements of  ${\bf V}_{\rm CKM}$ are non-zero,
the only significant off-diagonal elements are $ \rm |V_{us}|$ and
$ \rm |V_{cd}|$, giving the approximate matrix 
$$
{\bf V}_{\rm CKM} = \left(\matrix {\cos\theta & -\sin\theta&0\cr 
	\sin\theta&1-\cos\theta& 0\cr
	0 &0& 1\cr}\right),                   \eqno (5.6)
$$
where $\theta$, describing the mixing of generations 1 an 2,
is largest rotation angle given in equation (7.95) of [11].
Signs of off-diagonal elements are determined
by the requirement that columns of $ {\bf V}_{\rm CKM}$ are orthogonal.

Given that the CKM matrix is not simply a quark mixing matrix, but 
expresses a rotation in the C=$\pm 1$ coordinate systems
describing the three generations, equations (5.3) and (5.4) may also be
relevant to the leptons. In this case, maintaining the CKM relationship gives 
$$\eqalign{
\{{\rm e}^-:1\>1\>1\}_- &= \rm V_{ud}\{{\rm e}^-:1\>1\>1\}_+   + V_{us}\{\mu^-:1\>1\>\bar1\}_+ + V_{ub}\{{\tau}^-:1\>\bar1\>1\}_+\cr
	\{\mu^-:1\>1\>\bar1\}_-&=  \rm V_{cd}\{{\rm e}^-:1\>1\>1\}_+   + V_{cs}\{\mu^-:1\>1\>\bar1\}_+ + V_{cb}\{{\tau}^-:1\>\bar1\>1\}_+\cr
\{{\tau}^-:1\>\bar1\>1\}_-&= \rm V_{td}\{{\rm e}^-:1\>1\>1\}_+   + V_{ts}\{\mu^-:1\>1\>\bar1\}_+ + V_{tb}\{{\tau}^-:1\>\bar1\>1\}_+\cr}
\eqno (5.7)
$$ 
where the (CFG) quantum numbers for the leptons are taken from Table A.1 in the Appendix.
The corresponding relations for neutrinos are 
$$\eqalign{
\{\nu_e:\bar1\>1\>1\}_+  &=\rm V_{du}\{\nu_e:\bar1\>1\>1\}_- +V_{dc}\{\nu_\mu:\bar1\,1\,\bar1\}_- + V_{dt}\{{\nu}_\tau:\bar1\>\bar1\>1\}_-\cr
\{\nu_\mu:\bar1\,1\,\bar1\}_+&= \rm V_{su}\{\nu_e:\bar1\>1\>1\}_-+V_{sc}\{\nu_\mu:\bar1\,1\,\bar1\}_- + V_{st}\{{\nu}_\tau:\bar1\>\bar1\>1\}_-\cr
\{{\nu}_\tau:\bar1\>\bar1\>1\}_+&=\rm V_{bu}\{\nu_e:\bar1\>1\>1\}_- +V_{bc}\{\nu_\mu:\bar1\,1\,\bar1\}_- + V_{bt}\{{\nu}_\tau:\bar1\>\bar1\>1\}_-\cr}
\eqno (5.8)
$$
This could help to explain observed neutrino mixing process.

\beginsection \S6: Applications of the conservation law to fermion and meson interactions

The conservation law (QNC) was formulated in \S8 of [2] as an equation relating 
the quantum numbers associated with all the fermion lines linked to each node 
of Feynman diagrams: 
\vskip5pt
{\it All seven quantum numbers are conserved in fermion decays and interactions}
\vskip5pt
However, as noted in [3], applications of QNC need modification to take account 
of the dependence the quantum numbers F, G, which describe fermion generations, 
on the value of the parity quantum number C. Note, however, that C is not additive
like B,F and G, but multiplicative, and always takes the values $\pm 1$. 
The effect of the CKM transformation is to replace QNC with a modified (MQNC)
law that takes all generation changes into account. 

The `time' quantum number B has the effect of counting the 
number of fermions in an interactions, with anti-fermions counted as B=$-1$ .
Hence mesons, constructed from one fermion and one anti-fermion, all
have B=0.  QNC is not affected by the CKM transformation if all
the interacting fermions have either C=1, and only involve d,s,b,e$^-$,$\mu^-$,$\tau^-$, 
or C=$-1$, and only involve u,c,t,$\nu_e$,$\nu_\mu$,$\nu_\tau$. 
These classifications remain the same for the corresponding anti-fermions.
For example, positrons (e$^+$) are still C=+1 fermions, although C=$-1$.

Charged mesons are constructed from one C=1 fermion (or anti-fermion) 
and C=$-1$ fermion (or anti-fermion). In consequence, they are
described by a sum of two $\{\rm CFG\}$ triplets with opposite parities.
In order to apply MQNC it is necessary
to express both of these fermions in terms of either the C=1 or C=$-1$
coordinates using equations (5.3) or (5.4). Neutral mesons are constructed
either from two C=+1 or from two C=$-1$ fermions. It follows that
their most important decay is either into two C=+1 or two C$-1$ fermions.

Processes involving both C=1 or C=$-1$ fermions can satisfy 
MQNC if the equation holds separately for
fermions with each of the C values. An example is provided by  
the first generation $\beta$ decay $ {\rm d}\to {\rm u}+ {\rm e}^- + \bar\nu_e$, 
where d, e$^-$ are C=1 fermions and u, $\bar\nu_e$ are C$=-1$ fermions. Employing
the CDEFG values given in Table A.1, this decay has the BCDEFG quantum number equation
$$ 
\{{\rm d}_b:1\>1\>1\>1\>1\>1\}\to\{{\rm u}_b:1\>\bar1\>1\>1\>1\>1\} 
+\{{\rm e}^-:1\>1\>\bar1\>\bar1\>1\>1\}+\{\bar\nu_e:\bar1 \>1\>1\>1\>\bar1\>\bar1\}. \eqno (6.1)
$$
where the F,G quantum numbers cancel separately for both C=1 and C=$-1$ fermions. 
Different d,u quark colours change the signs of D and E in the same way, leaving
the equality unchanged. It follows that equation (6.1) satisfies QNC.

Given that equation (6.1) is independent of quark colour, and the sign of C
distinguishes fermions from anti-fermions, it is unnecessary to include the B, E 
and D quantum numbers explicitly. With this simplification (6.1) can be expressed in
terms of CFG quantum numbers as
$$ 
\{{\rm d}:1\>1\>1\}\to\{{\rm u}:\bar1\>1\>1\} 
+\{{\rm e}^-:1\>1\>1\}+\{\bar\nu_e:1\>\bar1\>\bar1\}.                       \eqno (6.1a)
$$
Analogous equations hold for second and third generation $\beta$ decays, viz.
$$\eqalign{ 
	\{{\rm s}:1\>1\>\bar1\}\to&\{{\rm c}:\bar1\>1\>\bar1\}
	+\{\mu^-:1\>1\>\bar1\}+\{\bar\nu_\mu:1\>\bar1\>1\},\cr 
	\{{\rm b}:1\>\bar1\>1\}\to&\{{\rm t}:\bar1\>\bar1\>1\} 
	+\{\tau^-:1\>\bar1\>1\}+\{\bar\nu_\tau:1\>1\>\bar1\}.}   \eqno (6.2)
$$
Scattering processes are described by moving one set of fermion quantum numbers 
to the other side of the above equations. For example, equation (6.1a) becomes 
$$ 
\{{\rm d}:1\>1\>1\}+\{\nu_e:\bar1\>1\>1\}
\to\{{\rm u}:\bar1\>1\>1\}+\{{\rm e}^-:1\>1\>1\},             \eqno (6.3)
$$
describing the scattering of the $\nu_e$ neutrino by a d quark to produce
a u quark and an electron. 

Meson quark/antiquark structures ensure that their quantum numbers 
A=B=D=E=0, restricting their descriptions to parity (C) and 
generation (defined by F,G). Their truncated meson {CFG} signatures, given 
in Table A.2 of the Appendix, are based on 
currently accepted quark/anti-quark descriptions of J=0 mesons.
However, as was pointed out in \S2, these simple
identifications of some observed neutral mesons are in doubt.

The quantum numbers obey QNC in interactions between neutrinos and massive fermions,
but apparently breakdown
This is currently explained in terms of intergenerational oscillations made possible
by their very small masses, 

Neutrino experiments all give average effects for large
numbers of events.  These show considerable mixing of 
generations of neutrinos that have travelled very large
distances. This is currently interpreted as
being due to neutrino oscillations in vacuo, referred 
to as the see-saw mechanism, and described by
the Pontecorvo-Maki-Nakagawa-Sakata 
(PMNS) matrix (e.g. [4]\S13.4,5 and [12] Chapter 1).  
This would produce changes in the quantum numbers F,G, and
is therefore not consistent with QNC or MQNC. Nevertheless, it
might be explained by the extremely small neutrino masses. Mechanisms 
that do satisfy QNC could also be significant, such as that
provided by neutral meson decays. Examples, in terms of CFG 
quantum numbers, are
$$\eqalign{
\rm \{D^0\equiv c\bar u:0\,0\,\bar 2 \}\to &\{\nu_\mu:\bar1\,1\,\bar1\} + \rm\{\bar\nu_e:1\,\bar1\,\bar1\},\cr
\rm \{T^0\equiv t\bar u:0\,\bar 2 \,0\} \to & \{\bar\nu_e:\>1\, \bar1\,\bar1\}+\rm \{\nu_\tau:\bar1\,\bar1\,1\},\cr 
\rm \{T^0_c\equiv t\bar c:0\,\bar 2\,2\}\to & \rm\{\nu_\tau:\bar1\,\bar1\,1\}+\{\bar\nu_\mu:1\,\bar1\,1\} .\cr }                                              \eqno (6.4)
$$ 
As all these decays relate C=$-1$ fermions, they satisfy QNC, but would be very 
difficult to observe. More easily observed interactions, involving charged mesons,
do not satisfy strict QNC as not all the fermions can have C=$-1$.  For example, ignoring
C distinctions of F,G, MQNC gives the dominant $\pi$ meson decays
$$
\{\pi^-:2\, 0\, 0\} \to \{\mu^-:1\, 1\,\bar1\} + \{\bar \nu_\mu:1\,\bar1\, 1\},
\>\{\pi^+:\bar2\,0\,0\} \to \{\mu^+:\bar1\, \bar1 \, 1\} + \{ \nu_\mu:\bar1 \, 1\, \bar1\},      \eqno (6.5)
$$
and
$$
\{\pi^-:2\, 0\, 0\} \to \rm \{e^-:1\, 1\,\bar1\} + \{\bar \nu_e:1\,\bar1\, 1\},
\>\{\pi^+:\bar2\,0\,0\} \to \rm \{e^+:\bar1\, \bar1 \, 1\} + \{ \nu_e:\bar1 \, 1\, \bar1\}.      \eqno (6.6)
$$
Observed decays into $\pi^- \to \rm e^-, \bar\nu_e$ and
$\pi^+ \to\rm e^+, \nu_e$ are far less frequent than the 
$\pi^\pm \to\mu^\pm $ decays (e.g. see [5] \S11.6).
A simple explanation of this could be that decays into $\mu^-$
release less free energy in the form of anti-neutrinos.  

QNC provides stronger constraints
than charge in meson decays. For example, the decays
$\rm D^+ \to K^- + 2\pi^+ $ and $\rm D^- \to K^- + 2\pi^- $
are observed, while the decays $\rm D^+ \to  K^+ +\pi^+ + \pi^-$ and
$\rm D^- \to K^- +\pi^+  + \pi^- $ are not observed. The
corresponding quantum number equations are
$$
\{\rm D^+\equiv c\bar d:\bar2 0 \bar2\} \to \{K^-:2 0 \bar2\} + 2\{\pi^+:\bar2 0 0\},\>\>\>\>
\{\rm D^-:2 0 2\} \to \{K^+:\bar2 0 2\} + 2\{\pi^-:2 0 0\}                       \eqno(6.7)
$$ 
and 
$$
\{\rm D^+:\bar2 0 \bar2\} \not \to \{K^+:\bar2 0 2\} + \{\pi^+:\bar2 0 0\} +  \pi^-\{2 0 0\},\>\>\>\>
\{\rm D^-:2 0 2\} \not \to  \{K^-:2 0 \bar2\}?  + \{\pi^+:\bar2 0 0\}+ \{\pi^-:2 0 0\}.\eqno(6.8)
$$
The significant difference is that G quantum numbers are conserved in equation (6.7) but 
not in equation (6.8). However, as $\rm D^-\{2 0 2\}$ and $\rm K^+\{\bar2 0 2\}$ 
have different values of C, equation (6.7) only gives the dominant decay.
It is also of interest to explore other decay products of $\{\rm D^+\equiv c\bar d\} $
as an example of charged meson decays. One approach is to use equation (5.3)
to express ${\rm c}:(\bar1\> 1 \>\bar1)_-$ in terms of SU(3)$_+$ quantum numbers, giving
$$\eqalign{
	\rm D^+&\equiv \{\rm c\bar d:(\bar1\> 1 \>\bar1)_- +(\bar1\> \bar1\> \bar1)_+ \} 
	\equiv {\rm c}:(\bar1\> 1 \>\bar1)_- + {\rm \bar d}:(\bar1\> \bar1\> \bar1)_+\cr 
	&=\rm\bar d:\{\bar1 \>\bar1 \>\bar1\}_+ +\rm V_{cd}\{{\rm d}:1\>1\>1\}_+ + V_{cs}\{{\rm s}:1\>1\>\bar1\}_+ + V_{cb} \{{\rm b}:1\>\bar1\>1\}_+ \cr 
	&=\rm (1-V_{cd}) \bar d\{\bar1\> \bar1\> \bar1 \}_+
	+ V_{cs}s:\{1\>1\>\bar1\}_+ + V_{cb} b:\{1\>\bar1\>1\}_+ ,} \eqno(6.9)                       
$$
where each of the three terms corresponds to a separate node in Feynman diagrams.
Similarly, $\rm D^+$ can be expressed in terms of SU(3)$_-$ quantum numbers 
by employing equation (5.4).

Neutral meson decays are dominated by their construction
of two quarks with the same parity, so that the CKM matrix 
is not involved. This is complicated, however, by the fact that
the observed mesons do not necessarily correspond to the simple
quark/anti-quark structures shown in Tables 2.1 and 2.2.
Another complication is that it is possible to use the
CKM matrix to express both quarks in terms opposite parities,
producing many other weak decays as described, for example,
in Chapter 14 of [5]. However, as this raises the controversial
question of parity non-conservation, a detailed analysis would
go beyond the aims of this work.

Interactions between neutral mesons 
listed either in Table 2.1 or in Table 2.2 are 
analogous to the interactions between gluons in quantum chromodynamics. 
Given that the algebraic description of SU(3) gluons is isomorphic
to that of the SU(3)$_-$ and SU(3)$_+$ neutral mesons, this is inevitable. 
It is, of course, consistent with QNC which, given the meson masses,
predicts, for example, the decays 
$$
\{\rm \bar T^0_c:02\bar2\}\to \rm \{D^0:0 0\bar2\} + \{\bar T^0:020\},\>\>
 \{\rm B^0_s:0\bar2 2\}\to\{ \bar K^0:0 0 2\} + \bar \{B^0 :0 \bar2 0\}.      \eqno(6.10)
$$ 
Changes of charged meson generations can also
be produced by interactions with neutral mesons, such as
$$
\rm\{ B^0:020\} + \{K^-:20\bar2\} =\{T_s^-:22\bar2\}.                   \eqno(6.11)
$$
  
\vskip 5pt

\beginsection \S7. Conclusions

Both the quark generations u,c,t, with parity $\rm C=-1$ and the quark generations d,s,b, 
with parity $\rm C=1$, are described by the Lie group SU(3). Tables
2.1 and 2.2, distinguish these groups as SU(3)$_-$ and SU(3)$_+$,
according to their quark parities. Comparison with the Gell-Mann matrix
representation of generators relates them to the neutral mesons.
Generators of the Lie group SU(6) $\equiv$  SU(3)$\otimes$U(1)$\otimes$SU(3)
are identified with J=0 mesons, comprising the
16 neutral generators of its SU(3)$_\pm$ subgroups, 
together with the 9 positively charged and 9 negatively charged 
generators listed in Table 3.1. 

The CKM matrix is shown to describe transformations between elements 
of SU(3)$_-$ and SU(3)$_+$. It is likely that
this transformation is determined by differences in magnitude of the C=1
and C=$-1$ meson masses. This possibility has recently been 
studied, for example in [13].

It is shown in \S6 how the quantum number Conservation Law (QNC), 
obtained in [2], can be modified to take into account the parity dependence 
of the F,G quantum numbers produced by the CKM matrix. A number of examples,
including inter-generational interactions involving mesons are given. It is
necessary to include the constraints of quantum number conservation in
computer analyses of experiments to get its practical benefits.

\beginsection Acknowledgement

I am grateful to Emeritus Professor Ron King for his assistance with the 
mathematical aspects of this work, especially for bringing reference [6] to my
attention.

\vfill\eject

\beginsection Appendix 

The following tables are limited to the quantum numbers of the first three (observed) 
generations. Fermion charges are given in [2] as
$$
\rm Q = Q_B + Q_C ={1\over 6}( D + E - BDE) - {1\over 2}( F + G - BFG)BC.        \eqno (A.1)
$$

$$\vbox
{\settabs 11 \columns
	\+ Table A.1& {(B=1) quantum numbers for the three observed generations of fundamental fermions}\cr 
	\+&|||||||||||||||||||||||||||||||||||||\cr
	\+&quark&C&D&E&F&G& $\rm Q_B$&$\rm Q_C$&Q\cr
	\+&|||||||||||||||||||||||||||||||||||||\cr
	\+&${\rm u}_b$    &$-1$   &$\>\>1$&$\>\>1$&$\>\>1$ &$\>\>1$  &1/6&1/2&$\>\>2/3$ \cr
	\+&${\rm u}_r$    &$-1$   &$\>\>1$&$-1$   &$\>\>1$ &$\>\>1$  &1/6&1/2&$\>\>2/3$ \cr
	\+&${\rm u}_g$    &$-1$   &$-1$   &$\>\>1$&$\>\>1$ &$\>\>1$  &1/6&1/2&$\>\>2/3$ \cr 
	\+ &&&\cr
	\+&${\rm d}_b$    &$\>\>1$&$\>\>1$&$\>\>1$&$\>\>1$ &$\>\>1$&1/6&$-1/2$&$-1/3$  \cr
	\+& ${\rm d}_r$   &$\>\>1$&$\>\>1$&$-1$   &$\>\>1$ &$\>\>1$&1/6&$-1/2$&$-1/3$  \cr
	\+&${\rm d}_g$    &$\>\>1$&$-1$   &$\>\>1$&$\>\>1$ &$\>\>1$&1/6&$-1/2$&$-1/3$  \cr 
	\+&|||||||||||||||||||||||||||||||||||||\cr
	\+&${\rm c}_b $       &$-1$   &$\>\>1$&$\>\>1$&$\>\>1$    &$-1$  &1/6&$1/2$&$\>\>2/3$\cr
	\+&${\rm c}_r $       &$-1$   &$\>\>1$&$-1$   &$\>\>1$    &$-1$  &1/6&$1/2$&$\>\>2/3$\cr
	\+& ${\rm c}_g $      &$-1$   &$-1$  &$\>\>1$ &$\>\>1$    &$-1$   &1/6&$1/2$&$\>\>2/3$\cr 
	\+ &&&\cr
	\+& ${\rm s}_b $      &$\>\>1$&$\>\>1$&$\>\>1$&$\>\>1$    &$-1$  &1/6&$-1/2$& $-1/3$  \cr
	\+&${\rm s}_r $       &$\>\>1$&$\>\>1$&$-1$   &$\>\>1$    &$-1$  &1/6&$-1/2$& $-1/3$ \cr
	\+& ${\rm s}_g $      &$\>\>1$&$-1$   &$\>\>1$&$\>\>1$    &$-1$  &1/6&$-1/2$& $-1/3$ \cr
	\+&|||||||||||||||||||||||||||||||||||||\cr		
	\+ &${\rm t}_b $  &$-1$   &$\>\>1$&$\>\>1$&$-1$  &$\>\>1$  &1/6&$1/2$&$\>\>2/3$ \cr
	\+ &${\rm t}_r $  &$-1$   &$\>\>1$&$-1$   &$-1$  &$\>\>1$   &1/6&$1/2$&$\>\>2/3$\cr
	\+ &${\rm t}_g $  &$-1$   &$-1$   &$\>\>1$&$-1$  &$\>\>1$   &1/6&$1/2$&$\>\>2/3$\cr 
	\+ &&&\cr
	\+ &${\rm b}_b $  &$\>\>1$&$\>\>1$&$\>\>1$&$-1$  &$\>\>1$   &1/6 &$-1/2$&$-1/3$ \cr	
	\+ &${\rm b}_r $  &$\>\>1$&$\>\>1$&$-1$   &$-1$  &$\>\>1$   &1/6&$-1/2$&$-1/3$  \cr	
	\+ &${\rm b}_g $  &$\>\>1$&$-1$   &$\>\>1$&$-1$  &$\>\>1$   &1/6&$-1/2$&$-1/3$  \cr
	\+&|||||||||||||||||||||||||||||||||||||\cr	  	
	\+ &$\nu_e $       &$-1$&$-1$   &$-1$   &$\>\>1$ &$\>\>1$ &$-1/2$&1/2&$\>\>\>\>0$\cr	
	\+ &$\nu_\mu $     &$-1$&$-1$   &$-1$   &$\>\>1$ &$-1$    &$-1/2$&1/2&$\>\>\>\>0$\cr	
	\+ &$\nu_\tau $    &$-1$&$-1$   &$-1$   &$-1$    &$\>\>1$ &$-1/2$&1/2&$\>\>\>\>0$\cr
	\+ &&&\cr
	\+ &e$^-$       &$\>\>1$&$-1$   &$-1$   &$\>\>1$ &$\>\>1$&$-1/2$&$-1/2$&$\>\>-1$ \cr
	\+ &$\mu^-$     &$\>\>1$&$-1$   &$-1$   &$\>\>1$ &$-1   $&$-1/2$&$-1/2$&$\>\>-1$ \cr
	\+ &$\tau^-$    &$\>\>1$&$-1$   &$-1$   &$-1$    &$\>\>1$&$-1/2$&$-1/2$&$\>\>-1$ \cr 
	\+&|||||||||||||||||||||||||||||||||||||\cr}
$$
\vfill\eject

The following table is based on J=0 meson structures given in [12], p.391 and [5], p.524. 
It distinguishes C dependent F,G quantum numbers as F$_\pm$ and G$_\pm$.
$$\vbox 
{\settabs 10 \columns\+  Table A.2: {CFG} signatures of the J=0 mesons based on their accepted quark structures\cr  
	\+|||||||||||||||||||||||||||||||||||||||||\cr
	\+q$\bar{\rm q}$ &C&$\>\>$F$_+$ &$\>\>$F$_-$& $\>\>$G$_+$& $\>\>$G$_-$  &meson&mass(Mev)\cr
	\+|||||||||||||||||||||||||||||||||||||||||\cr
	\+d$\bar{\rm u}$&$\>\>1$&1     &$-1$&$1$  &$-1$       &$\>\>\>\pi^-$&139.6\cr 
	\+u$\bar{\rm d}$&$\>\>1$&$-1$  &1   &$-1$ &1          &$\>\>\>\pi^+$&139.6\cr   
	\+s$\bar{\rm c}$&$\>\>1$&$1 $  &$-1$&$-1 $&$-1$       &    D$^-_s$&1968.5\cr 	    
	\+c$\bar{\rm s}$&$\>\>1$&$ -1$ &1   &$1$  &$1$        &$\>\>$D$^+_s$&1968.5\cr		           
	\+b$\bar{\rm t}$&$\>\>1$&$ -1$ &1   &$1$  & $-1$      &T$^-_b$&80385\cr 
	\+t$\bar{\rm b}$&$\>\>1$&$1 $  &$-1$&$-1$ &1          &$\>\>$ T$^+_b$&80385\cr 	
	\+&&&&&&\cr
	\+u$\bar{\rm c}$&$-1$&$\>\>0$  &$\>\>0$&0      &$\>\>2$&$\>\>\bar{\rm D}^0$&1864.5 \cr 
	\+c$\bar{\rm u}$&$-1$&$\>\>0$  &$\>\>0$&  0    &$-2$   &$\>\>$D$^0$        &1864.5\cr
	\+d$\bar{\rm s}$&$-1$&$\>\>0$  &$\>\>0$&$\>\>2$&0      &$\>\>$K$^0$        &497.6\cr           
	\+s$\bar{\rm d}$&$-1$&$\>\>0$  &$\>\>0$&$-2$   &0      &$\>\>\bar{\rm K}^0$ &497.6\cr 	
	\+&&&&&&\cr	
	\+u$\bar{\rm s}$&$\>\>1$&$ -1$  &1      &$1 $&$ 1$   &$\>\>$K$^+$&493.7\cr
	\+s$\bar{\rm u}$&$\>\>1$&$1 $   &$-1$   &$-1 $&$-1$  &$\>\>$K$^-$&493.7\cr
	\+c$\bar{\rm d}$&$\>\>1$&$-1$  &1      &$-1 $&$ -1$    &$\>\>$D$^+$&1869.3\cr 
	\+d$\bar{\rm c}$&$\>\>1$&$1 $   &$-1$   &$1 $&$1$    &$\>\>$D$^-$&1869.3\cr
	\+&&&&&&\cr	
	\+b$\bar{\rm d}$&$-1$&$-2$   &$\>\>0$    &$\>\>0$&$\>\>0$&$\>\>\bar{\rm B}^0$   &5279\cr
	\+d$\bar{\rm b}$&$-1$&$\>\>2$&$\>\>0$    &$\>\>0$&$\>\>0$&$\>\>$B$^0$             &5279\cr
	\+u$\bar{\rm t}$&$-1$&$\>\>0$&$\>\>2$    &$\>\>0$&$\>\>0$&$\>\>$$\bar{\rm T}^0$   &\cr 
	\+t$\bar{\rm u}$&$-1$&$\>\>0$&$-2$       &$\>\>0$&$\>\>0$&$\>\>$T$^0$             &\cr
	\+&&&&&&\cr		
	\+b$\bar{\rm u}$&$\>\>1$&$-1$     & $-1$&$\>\>1$   &$ -1$  &$\>\>$ B$^-$&5279\cr 
	\+u$\bar{\rm b}$&$\>\>1$   &$\>\>1 $ &$\>\>1$& $-1$   &$\>\>1$&$\>\>$ B$^+$&5279\cr 
	\+d$\bar{\rm t}$&$\>\>1$&$\>\>1 $ &$\>\>1$&$\>\>1 $&$-1$    &$\>\>$ T$^-$&\cr
	\+t$\bar{\rm d}$&$\>\>1$   &$-1$     & $-1$&$ -1$    &$\>\>1$ &$\>\>$ T$^+$&\cr
	\+&&&&&&\cr		
	\+t$\bar{\rm c}$&$-1$& $\>\>0$&$-2$   &$\>\>0$&$\>\>2$&$\>\>$ T$^0_c$          &\cr
	\+c$\bar{\rm t}$&$-1$&$\>\>0$ &$\>\>2$&$\>\>0$&$-2$   &$\>\>\>\>\bar{\rm T}^0_c$&\cr 
	\+b$\bar{\rm s}$&$-1$&$-2$    &$\>\>0$&$\>\>2$&$\>\>0$ &$\>\>\>\>\bar{\rm B}^0_s$&6368\cr 	
	\+s$\bar{\rm b}$&$-1$&$\>\>2$ &$\>\>0$&$-2$  &$\>\>0$ &$\>\>\>\>$B$^0_s$       &6368\cr 
	\+&&&&&&\cr
	\+t$\bar{\rm s}$&$\>\>1$   &$-1$  &$-1$   & $\>\>1$ &$\>\>\>\>1$&$\>\>$ T$^+_s $&\cr
	\+s$\bar{\rm t}$&$\>\>1$&$ 1$  &$\>\>1$&$-1$   &$-1$&$\>\>$ T$^-_s$&\cr 	
	\+b$\bar{\rm c}$&$\>\>1$&$- 1$ &$-1$   &$\>\>1$  &$\>\>-1$   &$\>\>$ B$^-_c$&6286 \cr     
	\+c$\bar{\rm b}$&$\>\>1$   &$1$   &$\>\>1$&$-1$  &$\>\>\>\>1$&$\>\>$ B$^+_c$&6286\cr	
	\+|||||||||||||||||||||||||||||||||||||||||\cr}
$$
Note that these identifications describe meson properties, and not
the internal dynamics that determine their masses.

\beginsection References

\frenchspacing

\item {[1]} Newman, Douglas (2021) Unified theory of elementary fermions and their interactions
based on Clifford algebras {arXiv:2108.08274}

\item{[2]} Newman, D.J. (2024) Quantum number conservation: a tool in the design and 
analysis of high energy experiments. J.Phys.G: Nucl. Part. Phys. 51 095002 

\item{[3]} Newman, D.J. (2024) Corrigendum: Quantum number conservation: a tool in the design and 
analysis of high energy experiments.(J.Phys.G: Nucl. Part. Phys. 51 095002 ) J.Phys.G: Nucl. Part. Phys. 52 019501

\item {[4]} Newman, Douglas (2022) Repairing the algebraic foundations of the 
Standard Model of particle physics {arXiv:2308.12295v5}

\item {[5]} Thomson, Mark (2013) Modern Particle Physics (Cambridge University Press)

\item {[6]} Hartanto, A. and Handoko, L.T. (2005) Grand Unified Theory based on the SU(6) symmetry.\vskip1pt
Phys.Rev.D71:095013 {arXiv:hep-ph/0504280}

\item {[7]} Deppisch, F. F., Hati, C., Patra, S., Sarkar,U. and Valle, J.W.F. (2016) 331 Models and 
Grand Unification: From Minimal SU(5) to Minimal SU(6). Phys.Lett.B 762, 432 {arXiv:1608.05334} 

\item {[8]} Cifcti, R., Cifcti, A.K. and Popov, O. (2022) Democratic parametrization and analysis for 331 model as a subgroup of SU(6). {arXiv:2212.00282}

\item {[9]} Cifcti, R., Cifcti, A.K. and Popov, O. (2024) New parametrization and analysis for E6 inspired 331
model. DOI: 10.31526/LHEP-481  {arXiv:2211.16529}

\item {[10]} Angelescu,A., Bally, A., Blasi,S. and Goertz, F. (2022) Minimum SU(6) Gauge-Higgs Grand Unification.
Phys.Rev.D105.035026 {arXiv:2104.07366}

\item {[11]} Bettini, Alessandro (2008) Introduction to Elementary Particle Physics (Cambridge University Press)

\item{[12]} Giunti, Carlo and Kim, Chung W. (2007) Fundamentals of Neutrino Physics and
Astrophysics (Oxford University Press)

\item {[13]} Patel, Aditya Ankur and Singh, Tejinder P. CKM matrix parameters from the exceptional Jordan algebra. {arXiv:2305.00668v2 }

\end